\documentclass[journal,transmag]{IEEEtran}
\ifCLASSINFOpdf
\else
\fi
\usepackage{breqn}
\usepackage[noadjust]{cite}
\usepackage{xcolor}

\usepackage{graphicx}
\usepackage{amsmath, amsfonts}
\usepackage{amsmath, amssymb}
\usepackage{graphicx}
\usepackage{rotating}
\usepackage{multirow}
\usepackage{array}
\usepackage{xcolor}
\usepackage{url}
\usepackage{pgfplots}
\usepackage{verbatim}
\usepackage{tikz}
\usepackage{algorithm}
\usepackage{algorithmic}
\usepackage{breqn}
\usepackage{amsmath}
\usepackage[utf8]{inputenc}
\usepackage{nomencl}
\usepackage{array}
\usepackage{aurical}
\usepackage[T1]{fontenc}
\usepackage{soul}

\newcommand{\tabmore}{\hspace{10pt}}

\usetikzlibrary{mindmap,shadows,backgrounds}
\graphicspath{{figures/}}
\pgfplotsset{compat=1.14}

\begin{document}
%
\title{Transfer Reinforcement Learning for 5G-NR mm-Wave Networks}


\author{\IEEEauthorblockN{Medhat Elsayed\IEEEauthorrefmark{1}, ~\IEEEmembership{Student Member,~IEEE}
Melike Erol-Kantarci\IEEEauthorrefmark{1}, ~\IEEEmembership{Senior Member,~IEEE} \\ and
Halim Yanikomeroglu\IEEEauthorrefmark{2}, ~\IEEEmembership{Fellow,~IEEE}}
\IEEEauthorblockA{\IEEEauthorrefmark{1}School of Electrical Engineering and Computer Science,
University of Ottawa, Ottawa, Canada}
\IEEEauthorblockA{\IEEEauthorrefmark{2}Department of Systems and Computer Engineering, Carleton University, ON, Canada}}

\markboth{Accepted to IEEE Transactions On Wireless Communications, December~2020}%
{Medhat \MakeLowercase{\textit{et al.}}: IEEE Transactions On Wireless Communications}


%



\IEEEtitleabstractindextext{%
\begin{abstract}
In this paper, we aim at interference mitigation in 5G millimeter-Wave (mm-Wave) communications by employing beamforming and Non-Orthogonal Multiple Access (NOMA) techniques with the aim of improving network's aggregate rate. Despite the potential capacity gains of mm-Wave and NOMA, many technical challenges might hinder that performance gain. In particular, the performance of Successive Interference Cancellation (SIC) diminishes rapidly as the number of users increases per beam, which leads to higher intra-beam interference. Furthermore, intersection regions between adjacent cells give rise to inter-beam inter-cell interference. To mitigate both interference levels, optimal selection of the number of beams in addition to best allocation of users to those beams is essential. In this paper, we address the problem of joint user-cell association and selection of number of beams for the purpose of maximizing the aggregate network capacity. We propose three machine learning-based algorithms; transfer Q-learning (TQL), Q-learning, and Best SINR association with Density-based Spatial Clustering of Applications with Noise (BSDC) algorithms and compare their performance under different scenarios. Under mobility, TQL and Q-learning demonstrate $12\%$ rate improvement over BSDC at the highest offered traffic load. For stationary scenarios, Q-learning and BSDC outperform TQL, however TQL achieves about $29\%$ convergence speedup compared to Q-learning.
\end{abstract}

}

\maketitle

\IEEEdisplaynontitleabstractindextext

\IEEEpeerreviewmaketitle

\section{Introduction}
\IEEEPARstart{N}{ext}-generation wireless networks are expected to carry heterogeneous traffic loads with high Quality of Service (QoS) expectations including high capacity, low latency and enhanced reliability \cite{ITUR, DBLP, ramona}. The recent availability of the millimeter-Wave (mm-Wave) band between $30$ and $300$ GHz is a promising solution for spectrum scarcity in the next-generation wireless networks, where wide bandwidth can be provided for high data rate services. Indoor environments such as schools, hospitals, and shops, as well as outdoor environments such as parks, and city centres are examples of regions of mm-Wave support \cite{7010535}. However, the coverage of mm-Wave systems is limited due to the poor propagation characteristics of mm-Wave signals and their sensitivity to blockages such as buildings and people. In order to overcome such signal degradation, mm-Wave systems utilize directional communication through a large number of antennas (i.e., they use beamforming \cite{7959867}). In addition, power domain Non-Orthogonal Multiple Access (NOMA) provides opportunities to increase the spectral efficiency of wireless networks by superposing signals of multiple users on the same time and frequency resources while allocating different power levels to those signals \cite{6692652, 8647935, 8352621}. In consequence, Successive Interference Cancellation (SIC) technique is needed at the receiver side to demodulate respective users' signals \cite{7263349}. 

Despite the capacity gains promised by integrating mm-Wave, beamforming and NOMA, several technical challenges must be overcome, one of them being interference related performance degradation. In particular, intra-beam interference and inter-cell interference hinder such promised capacity gains. With NOMA, users' signals are superposed on the same time/frequency resources with different power levels. In turn, this incurs intra-beam interference which degrades the decoding performance of SIC technique (i.e., decoding performance of SIC diminishes rapidly as the number of users per beam increases \cite{6861434}). Furthermore, inter-cell interference arises due to intersection among beams that belong to different cells. Consequently, balancing the number of users covered by different beams is needed to maintain high performance of SIC. In order to accomplish this, we propose a joint user-cell association and number of beams selection for sum rate maximization in a fifth-Generation (5G) mm-Wave network. 

Several works in the literature have addressed the problem of sum rate maximization in mm-Wave networks. For example, in \cite{8454272}, the authors address inter-cluster and intra-cluster interference in a mm-Wave network for sum rate maximization through users clustering and NOMA power allocation. With beamforming, adjacent users tend to have correlated channel characteristics. As such, the authors propose a K-means algorithm that cluster users according to their channel features. Furthermore, they derive optimal NOMA power allocation policy in a closed form. With the aid of coalitional game theory, authors in \cite{8684765} propose a low complexity algorithm for users clustering in a single-cell mm-Wave system with the aim to maximize sum rate of the system. An optimal power allocation within each cluster has been proposed thereafter. 

The previous works differ with our work in three aspects. First, the previous works consider a single-cell scenario that aims to solve the clustering and power allocation problem. This corresponds to a centralized approach. In this work, we consider a multi-cell scenario where each cell acts as an independent agent (i.e., multi-agent scenario) that aims to mitigate interference by solving the joint user-cell association and number of beams selection. Second, the previous works employ a closed-form optimization technique which is adding a prohibitive complexity in implementation. Third, we propose three machine learning-based algorithms and analyze their performance based on the network scenario with different user deployments and under mobility. More specifically, a transfer reinforcement learning technique is proposed, where knowledge from an expert's task is transferred to a learner's task. This helps in utilizing samples of experience efficiently, hence speeding up the convergence of the learner task. To the best of our knowledge, this is the first time that transfer reinforcement learning is employed to address interference mitigation in a mm-Wave network with beamforming and NOMA.

The paper is organized as follows. Section \ref{sec:relatedwork} provides a summary of the recent work that uses transfer learning in enhancing the performance of wireless networks. Besides covering the basics of reinforcement learning in section \ref{sec:background}, we introduce the fundamentals of transfer reinforcement learning, with a focus on the adopted methodology, namely Transfer via Inter-Task Mapping (TvITM). Afterwards, Section \ref{sec:sysmodel} introduces our system model. Section \ref{sec:transferRL} presents the three proposed algorithms for joint user-cell association and number of beams selection. Simulation settings and performance evaluation is presented in section \ref{sec:perfEval}. Finally, Section \ref{sec:conclusion} concludes the paper. 

\section{Related Work}\label{sec:relatedwork}
An increasing number of efforts in the literature have been devoted to applying machine learning, and specifically reinforcement learning, to radio resource management in wireless networks \cite{8758918, 8714026, 8382166, 8755300, 7429691}. User-cell association has been addressed using reinforcement learning in several works. In \cite{8834857}, authors introduced a vehicle-road side unit association using deep reinforcement learning and asynchronous actor-critic in a mm-Wave network. The algorithm aims at maximizing the time average rate per vehicle while ensuring a target minimum rate for vehicles with low signaling overhead. In \cite{8714042}, authors propose a distributed Q-learning algorithm for user-cell association. In particular, Q-learning is employed to perform cell-level and user-level optimizations, where the algorithm aims to determine the best cell range extension offsets at the cell-level and the best weights of each user for efficient user-cell association at the user-level. The algorithm shows better results in terms of user satisfaction, outage, and minimization of dissatisfaction when satisfaction is not attained. In \cite{9058982}, a symbiotic radio network is considered, where an Internet-of-Things (IoT) network parasitizes in a primary network. In particular, the objective is to maximize the sum rate of IoT devices by associating each IoT user to one cellular user for information transmission. To achieve this association, the authors proposed two deep reinforcement learning algorithms, in which centralized and distributed approaches were conducted. In \cite{9145209}, authors address the problem of joint user-cell association and spectrum allocation with the aim to achieve fairness among users. This is achieved by addressing load balancing among cells by employing an online deep reinforcement learning algorithm. In particular, multiple parallel neural networks are used to output the user association decisions. The neural networks are trained from a shared memory that stores the best association decision in an online setup (i.e., no need for labeled data as training data is drawn from reinforcement learning's experience). The algorithm outperforms the greedy algorithm.

Recent attention is directed toward transfer learning, where knowledge is transferred from a source task to a target task. Authors in \cite{8917592} address the problem of minimizing network-wide power consumption subject to reliability, where a joint power and resource allocation for Ultra-Reliable Low-Latency Communication (URLLC) in vehicular networks is proposed. Using extreme value theory, extreme events (i.e., events with queue length exceeding a predefined threshold) are learned via a distributed approach based on federated learning. Furthermore, the proposed approach accounts for delays incurred by information transfer over wireless links. The proposed solution reveals an accuracy that is close to the centralized solution while significantly reducing the amount of information exchange. 

Besides federated learning, imitation learning has been adopted in a few works. In \cite{8891075}, the authors introduce an imitation learning technique to address the resource allocation problem in Device-to-Device (D2D) communications with the objective to maximize the minimum data rates of D2D pairs. Typically, the resource allocation problem can be formulated as a Mixed Integer Nonlinear Programming (MINLP) and possibly solved using a branch-and-bound algorithm. With imitation learning, the authors aim at learning a good auxiliary prune policy for accelerating the branch-and-bound algorithm. Simulation results demonstrate the capability of imitation learning in achieving optimality with reduced computational complexity.

A similar effort in \cite{8761327} used transfer learning via self-imitation to find a near-optimal solution for MINLP resource management in Cloud-Radio Access Networks (RANs) with the objective to minimize network power consumption. Again, the proposed framework aims to accelerate the branch-and-bound algorithm. Furthermore, it addresses the problem of task mismatch which occurs when the network settings change (i.e., test setting is different than training setting). 

Transfer actor-critic reinforcement learning is proposed in \cite{6747280}, which is the closest work to our contribution. The authors start by developing a reinforcement learning framework to control the switching operation of base stations through a centralized controller with the objective to minimize energy consumption in the network. Furthermore, they extend the framework with a transfer actor-critic algorithm that transfers the learned knowledge at a source controller to a target controller. Results demonstrate the rapid convergence of the transfer actor-critic algorithm. Unlike \cite{6747280}, our work has two main differences. Our work considers transfer learning between different tasks, where the source task is user-cell association and the target task is joint user-cell association and number of beams selection. This is unlike the work \cite{6747280}, where the knowledge difference lies in the dynamicity of the traffic model. Furthermore, our proposed algorithm aims to transfer knowledge from a simple task to a more complex task (i.e., joint decisions are harder as it requires correlation between the different parameters with respect to their impacts on environment changes). 

Transfer reinforcement learning with echo state neural network has been proposed in \cite{8648419} for resource allocation for wireless virtual reality networks. In particular, small-cell base stations aim at performing efficient uplink and downlink resource block allocation so as to maximize users' successful transmission probabilities. To achieve that, base stations utilize the correlation between users' uplink tracking information and downlink content data.

Our previous work in \cite{icc2020} presented a Q-learning algorithm to address the problem of interference mitigation in mm-Wave networks where joint user-cell association and inter-beam power allocation was considered for sum rate maximization. In this paper, we propose algorithms that learn the efficient number of beams formed by each next-generation NodeB (gNB). As such, we perform joint user-cell association and number of beams selection for sum rate maximization. In addition, here, we propose transfer reinforcement learning and evaluate its performance with respect to other algorithms with stationary and mobile users. 

\section{Background}\label{sec:background}

\subsection{Reinforcement Learning}
The problem of reinforcement learning is a straightforward framing of learning from interaction between a decision-maker (i.e., an agent) and its environment \cite{Sutton1998}. The agent observes a state that characterizes its environment. The state of the environment should compactly retain relevant information of the environment through immediate and past sensations. A state signal that satisfies this condition is said to have Markov property. Therefore, the reinforcement learning can be cast as a Markov Decision Process (MDP) with the four-element tuple: \{states, actions, transition probabilities, and reward function\}. In particular, at each time step $\tau$, the agent receives some representation of the environment's state $S_{\tau} \in \text{\Fontauri\bfseries S}$, where $\text{\Fontauri\bfseries S}$ is the set of possible states. Afterwards, the agent selects an action $A_{\tau} \in \text{\Fontauri\bfseries A}(S_{\tau})$, where $\text{\Fontauri\bfseries A}(S_{\tau})$ is the set of possible actions available in state $S_{\tau}$. At the next time step $(\tau+1)$, the agent receives a reward value $R_{\tau+1}$ in response to the taken action and the environment's state changes to $S_{\tau+1}$. Furthermore, the transition probabilities, $p(s'|s, a) = Pr\{S_{\tau+1}=s'|S_{\tau} =s, A_{\tau} =a\}$, defines the probability that the environment's state changes from $S_{\tau} = s$ to $S_{\tau+1}=s'$ when the agent performs action $A_{\tau} = a$. 

The ultimate goal of a reinforcement learning's agent is to identify the best policy that maximizes its total expected reward as follows: 
\begin{equation}
    \max_{\pi(s)} \mathbb{E}[R_{\tau+1} + \gamma R_{\tau+2} + \gamma^2 R_{\tau+3}... | S_{\tau} = s],
\end{equation}
where $0 \leq \gamma \leq 1$ is a discount factor that reduces the contribution of future rewards in addition to maintaining a stability in computations. $\pi(s)$ is a policy that defines the optimal action at state $s$. In particular, the mapping from state $S_{\tau}$ to action $A_{\tau}$ is performed by following a stochastic policy $\pi(a|s) = Pr\{A_{\tau} = a|S_{\tau} = s\}$. As such, the goal of a reinforcement learning agent is to seek a policy that maximizes its total expected discounted reward over the long run. To achieve that, a value function is used to quantify how good is a certain policy given a state-action pair. An action-value function can be defined as follows:
\begin{equation}
    q_{\pi}(s, a) = \mathbb{E}_{\pi}[R_{\tau+1} + \gamma R_{\tau+2} + \gamma^2 R_{\tau+3}... | S_{\tau} = s, A_{\tau} = a],
\end{equation}
where $q_{\pi}(s, a)$ is the action-value (i.e., quality value or Q-value), of policy $\pi$ when starting at $s^{th}$ state and taking $a^{th}$ action. The optimal value function can be computed through a brute-force method which becomes intractable for large state-action space. Instead, Q-learning, a model-free reinforcement learning algorithm, is used to iteratively approximate the Q-values. Q-learning is a temporal difference method that uses the following update rule to approximate an agent's policy:
\begin{dmath}
    q_{\tau}(S_{\tau}, A_{\tau}) \gets q_{\tau}(S_{\tau}, A_{\tau}) + \alpha [R_{\tau+1} + \gamma \max\limits_{A_{\tau+1}} q_{\tau+1}(S_{\tau+1}, A_{\tau+1}) - q_{\tau}(S_{\tau}, A_{\tau})],
    \label{eq:qvalue}
\end{dmath}
where $\max\limits_{A_{\tau+1}} q_{\tau+1}(S_{\tau+1}, A_{\tau+1})$ computes an approximate of the Q-value at the next state $S_{\tau+1}$. 

\subsection{Transfer Reinforcement Learning}
The idea behind transfer learning is to exploit the knowledge learned about one task to improve generalization in another task \cite{3086952}. Indeed, humans can re-utilize their learned knowledge from a previous task in solving new tasks more rapidly or with better solutions \cite{5288526}. This reduces the need for a large number of training samples, which is a common problem in reinforcement learning. For example, Temporal Difference (TD) methods, such as Q-learning, suffer from slow convergence due to the need of a large number of training samples of experience, commonly collected via trial-and-error approach over large number of iterations \cite{playingAtari}. Hence, the key motivation to transfer learning in TD methods is to reduce the amount of samples needed for learning the target task, and to reduce the convergence time. Furthermore, deep reinforcement learning can be considered as another technique to improve convergence, where efficient representations of the environment are drawn from high-dimensional input data that are further used to generalize over past experiences \cite{DQN}. However, deep reinforcement learning generalizes over a localized domain (i.e., same knowledge domain), whereas transfer reinforcement learning aims at transferring knowledge across domains.

\begin{figure}
    \centering
    \includegraphics[scale=0.65]{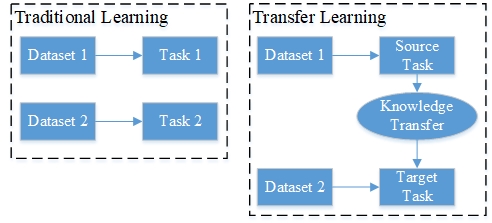}
    \caption{Conceptual explanation of the difference between traditional and transfer reinforcement learning.}
    \label{fig:transferLearning}
\end{figure}

\begin{figure}
    \centering
    \includegraphics[scale=0.7]{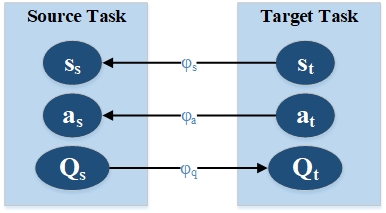}
    \caption{Transfer via Inter-Task Mapping.}
    \label{fig:TvITM}
\end{figure}

Fig. \ref{fig:transferLearning} presents a conceptual comparison between traditional and transfer reinforcement learning approaches. In particular, transfer learning considers knowledge transfer across tasks. Such knowledge transfer can transcend fixed or different domains. With fixed domain, the state-action spaces of the source and the target task are equivalent, whereas the objective, represented by the reward function, might differ. Transfer with different domains constitute different state-action spaces for source and target tasks \cite{crossDomainTRL, crossDomainTRL2}. 
In here, we adopt the TvITM approach proposed in \cite{JMLR07taylor}, where we consider transfer occurring from a single source task to a single target task. Fig. \ref{fig:TvITM} presents a conceptual model for TvITM approach. In particular, the expert's reinforcement learning task is defined as an MDP with the four-element tuple $\{s_s, a_s, T_s, r_s\}$, where $s_s$ is the state, $a_s$ is the action, $T_s$ is the transition function, and $r_s$ is the reward function of the source (expert) task. Similarly, the learner's task is defined as an MDP with the tuple $\{s_t, a_t, T_t, r_t\}$. While the state-action space defines the domain, the transition and reward functions define the objective of the task. As such, if both source and target tasks have the same state-action space, the transfer is said to be across fixed domain, otherwise it is a transfer across different domains. TvITM works as shown in Fig. \ref{fig:TvITM}. The state-action pair of the target, $(s_t, a_t)$, are mapped to the state-action pair of the source, $(s_s, a_s)$, via a state and action mapping functions, $\phi_s$ and $\phi_a$, respectively. Afterwards, the Q-value, $Q_s$, corresponding to $(s_t, a_t)$ is retrieved from the Q-table of the source task and mapped to a Q-value of the target task, $Q_t$, via a mapping function $\phi_q$. 

\section{System Model}\label{sec:sysmodel}
\textit{\textbf{Notations:}} In the remainder of this paper, bold face lower case characters denote column vectors, while non-bold characters denote scalar values. The operators $(.)^T$, $(.)^H$ and $|.|$ correspond to the transpose, the Hermitian transpose, and the absolute value, respectively.

\begin{figure}
    \centering
    \includegraphics[scale=0.37]{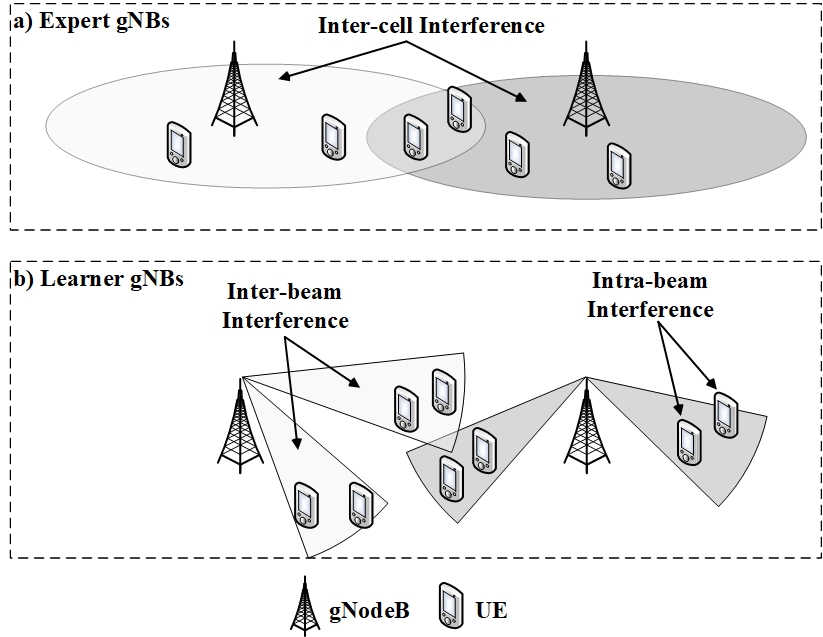}
    \caption{Network model of transfer learning in reinforcement learning with two expert and two learner gNBs.}
    \label{fig:networkModel}
\end{figure}

Consider a downlink mm-Wave-NOMA system with $e \in \text{\Fontauri\bfseries E}$ and $l \in \text{\Fontauri\bfseries L}$ expert and learner gNBs respectively as shown in Fig. \ref{fig:networkModel}. It is worth noting that expert and learner gNBs are spatially separated with no intersection zones. More specifically, Fig. \ref{fig:networkModel}a and Fig. \ref{fig:networkModel}b are used in conjunction when applying transfer reinforcement learning, whereas Fig. \ref{fig:networkModel}b is only used in case of Q-learning and Best SINR association with Density-based Spatial Clustering of Applications with Noise (BSDC). In addition, we consider two scenarios for users deployment. The first scenario considers stationary users, where initial positions follow Poisson Cluster Process (PCP). In PCP, the parent process follows a uniform distribution and the users of a cluster are uniformly deployed within a circular disk around the cluster center \cite{8454272}. The second scenario considers random waypoint mobility, where initial positions of the users follow PCP distribution. 

Expert and learner gNBs are each equipped with $M \in$ {\Fontauri\bfseries M} uniform linear array antennas to communicate with its associated single-antenna UEs, which constitute a Multiple Input Single Output (MISO) scenario. In addition, downlink NOMA power allocation is used to multiplex messages of UEs in the power domain (i.e., allocating different power levels to signals of different UEs). Consequently, UEs should employ SIC technique to demodulate their respective signals. It is worth mentioning that expert gNBs utilizes a single beam to communicate with its associated users. However, learner gNBs use a clustering algorithm to group User Equipments (UEs) that can be covered by a single beam, forming up to $k \in$ {\Fontauri\bfseries K} beams. Henceforth, we use cluster and beam interchangeably. Within each beam, downlink NOMA power allocation is used to multiplex messages of UEs in the power domain, and UEs use SIC technique at the receiver side. Furthermore, we consider that learner gNBs use K-means clustering algorithm and closed-form NOMA power allocation as proposed in \cite{8454272}. In particular, the k-means algorithm clusters UEs according to the correlation of their wireless channel properties (i.e., users with correlated channels are more likely to be located close to each other). 

In this work, we assume a mm-Wave channel with Line-of-Sight (LoS) link, hence the gain of the LoS path is significantly larger than the gain of the Non-LoS (NLoS) path \cite{7279196}. As such, the mm-Wave channel between an expert gNB and its associated UE can be considered as a single-path mm-Wave channel. It is worth mentioning that machine learning algorithm may benefit from such simple channel model. However, the idea of transfer learning is still applicable which is the main contribution of this paper. We plan to extend this work in the future with more realistic channel models. The mm-Wave channel is defined as follows:
\begin{equation}
    \boldsymbol{h}_{k,u} = \boldsymbol{v}(\theta_{k,u}) \frac{\alpha_{k,u}}{\sqrt{L}(1 + d_{k,u}^{\eta})},
\end{equation}
where $\boldsymbol{h}_{k,u} \in \mathbb{C}^{M \times 1}$ is the channel complex coefficient vector of $u^{th}$ UE and $k^{th}$ beam (i.e., link $(k,u)$), $\alpha_{k,u} \in \mathbb{C}N(0, \sigma^2)$ is the complex gain, and $d_{u}^{\eta}$ is the distance of $(k,u)^{th}$ link with pathloss exponent $\eta$. In addition, $\boldsymbol{v}(\theta_{k,u})$ is the steering vector, which is represented as follows:
\begin{equation}
    \boldsymbol{v}(\theta_{k,u}) = [1, e^{-j 2 \pi \frac{D}{\lambda} \sin(\theta_{k,u})}, ..., e^{-j 2 \pi (M-1) \frac{D}{\lambda} \sin(\theta_{k,u})}]^T,
\end{equation}
where $D$ is the gNB's antenna spacing, $\lambda$ is the wavelength, and $\theta_{k,u}$ is the Angle of Departure (AoD).

\subsection{Expert gNB: Only for Transfer Reinforcement Learning}
In this work, we employ the Q-learning algorithm for expert gNBs for sum rate maximization. In particular, expert gNBs aims at improving the sum rate through user-cell association. Sum rate can be modeled as follows:
\begin{equation}
    C_{e} =  \sum\limits_{e \in \text{\Fontauri\bfseries E}}  \sum\limits_{u \in \text{\Fontauri\bfseries U}_e} \omega \log_2(1 + \Gamma_{u,e}),
    \label{eq:sumRateExpert}
\end{equation}
where $\omega$ is the bandwidth, $\text{\Fontauri\bfseries U}_e$ is the set of UEs covered by $e^{th}$ expert gNB, and $\Gamma_{u,e}$ is the SINR of $(u,e)^{th}$ link, which can be expressed as
\begin{equation}
    \Gamma_{u,e} = \frac{P_{e} \beta_{u,e} |\boldsymbol{h}_{u,e}|^2}{ \sum\limits_{\substack{m \in \text{\Fontauri\bfseries E} \\ m \neq e}} P_{m} |\boldsymbol{h}_{u,m}|^2 + \sigma^2},
    \label{eq:sinrExpert}
\end{equation}
where $P_{e}$ denotes the power of the $e^{th}$ expert gNB, and $\beta_{u,e}$ is the NOMA power allocation factor of $(u,e)^{th}$ link. $\boldsymbol{h}_{u,m}$ represents the channel vector between the $m^{th}$ interfering expert gNB and $u^{th}$ UE, $P_m$ is the power of the $m^{th}$ interfering expert gNB, and $\sigma^2$ represents receiver's noise variance. 

\subsection{Learner gNBs} 
The objective of learner gNBs is equivalent to expert gNBs' objective, which is improving the sum rate of the network. However, learner gNBs aim to accomplish this through joint user-cell association and selection of the number of beams. In particular, sum rate of learner gNBs can be calculated as follows: 
\begin{equation}
    C_{l} = \sum\limits_{l \in \text{\Fontauri\bfseries L}} \sum\limits_{k \in \text{\Fontauri\bfseries K}_l} \sum\limits_{u \in \text{\Fontauri\bfseries U}_k} \omega \log_2(1 + \Gamma_{k,u,l}),
    \label{eq:sumRate}
\end{equation}
where $\text{\Fontauri\bfseries K}_l$ is the set of beams formed by $l^{th}$ gNB, and $\text{\Fontauri\bfseries U}_k$ is the set of UEs covered by $k^{th}$ beam. $\Gamma_{k,u,l}$ is the SINR of $(u,k,l)^{th}$ link, which can be expressed as:
\begin{equation}
    \Gamma_{k,u,l} = \frac{P_{k,l} \beta_{k,u,l} |\boldsymbol{h}_{k,u,l}^H \boldsymbol{w}_{k,l}|^2}{I_1 + I_2 + \sigma^2},
    \label{eq:sinr}
\end{equation}
\begin{equation}
    I_1 = P_{k,l} |\boldsymbol{h}_{k,u,l}^H \boldsymbol{w}_{k,l}|^2 \sum\limits_{\substack{i \neq u \\ O(i) > O(u)}} \beta_{k,i,l},
    \label{eq:inter1}
\end{equation}
\begin{equation}
    I_2 = \sum\limits_{l \in \text{\Fontauri\bfseries L}} \sum\limits_{\substack{m \in \text{\Fontauri\bfseries K}_l \\ m \neq k}} P_m |\boldsymbol{h}_{m,u,l}^H \boldsymbol{w}_{m,l}|^2,
    \label{eq:inter2}
\end{equation}
where $P_{k,l}$ denotes the power allocated to $k^{th}$ beam of $l^{th}$ gNB, $\beta_{k,u,l}$ is the power allocation factor of $(k,u,l)^{th}$ link, $\boldsymbol{w}_{k,l}$ is the beamforming vector. $I_1$ in (\ref{eq:inter1}) represents intra-beam interference caused by NOMA power allocation (i.e., UEs under the same beam share the same time/frequency resources). In addition, $I_2$ in (\ref{eq:inter2}) represents inter-beam interference (i.e., inter- or intra-cell inter-beam interference). $O(u)$ denotes the decoding order of $u^{th}$ user. Finally, $\boldsymbol{h}_{m,u,l}$ represents the channel vector between the $m^{th}$ interfering beam and $u^{th}$ user.

\section{Proposed Machine Learning Algorithms}\label{sec:transferRL}

\subsection{Transfer Reinforcement Learning}

\subsubsection{Expert: Q-Learning}\label{sec:expertQlearning}
Conventional Q-learning has been adopted for the expert gNBs. In particular, the state, $s_{e}$, of $e^{th}$ expert gNB is formulated to capture the level of interference represented as follows:
\begin{equation}
    s_{e} = 
    \begin{cases}
        s_0, \tabmore \overline{\Gamma}_e \geq \Gamma_{th}, \\
        s_1, \tabmore \text{otherwise},
    \end{cases}
    \label{eq:states}
\end{equation}
where $\Gamma_{th}$ is the SINR's threshold for successful packet decoding, and $\overline{\Gamma}_e$ is the average SINR of $e^{th}$ expert gNB due to downlink transmission to its associated users, which can be formulated as follows:   
\begin{equation}
    \overline{\Gamma}_e = \frac{1}{(U)} \sum\limits_{u \in \text{\Fontauri\bfseries U}_e} \Gamma_{u,e}
\end{equation}
Where, $\Gamma_{u,e}$ is the SINR of $u^{th}$ user associated to $e^{th}$ expert gNB. The state $s_0$ is visited whenever the achieved average SINR meets the minimum threshold, and $s_1$ is visited otherwise. The expert's actions are the user-cell association decision which is formulated as $\boldsymbol{a}_e = [\boldsymbol{\delta}_{u,e}; u \in \text{\Fontauri\bfseries U}_e, e \in \text{\Fontauri\bfseries E}]$, where $\boldsymbol{\delta}_{u,e}$ represents a logical indicator of $u^{th}$ UE's association to $e^{th}$ gNB. The reward function of $e^{th}$ gNB, $r_e$, is formulated using a sigmoid function as follows:
\begin{equation}
    r_e = \frac{1}{1 + e^{-0.5 (\overline{\Gamma}_e - 0.5 \Gamma_{th})}}.
    \label{eq:reward}
\end{equation}
Eq. (16) implies that a better SINR value, $\overline{\Gamma}_e$, rewards the expert gNB with higher reward value. 

\subsubsection{Learner: Transfer Q-learning (TQL)}
The learner gNB employs a TQL approach which is based on the framework of transfer via inter-task mapping in \cite{JMLR07taylor}. In particular, the ultimate goal of TQL's agent is to speed up its learning process in a target task by mapping a learned value function (i.e., Q-value function) of a different but related source task. In TQL, the target task is performed by the learner gNB (i.e., joint user-cell association and selection of number of beams), whereas the source task is performed by the expert gNB (i.e., user-cell association). In addition, we assume that knowledge of expert gNB (i.e., converged Q-table) becomes available to learner gNB before the latter starts its learning process.

Therefore, the formulation of TQL is similar to conventional Q-learning with the addition of a mapping function. In particular, the mapping function is used to import a Q-value from the expert gNB's knowledge domain. Such Q-value acts as signal to guide, and speed up, the Q-learning algorithm of the learner gNB. The following lines explain the MDP tuples of TQL.

\begin{itemize}
    \item \textbf{Agents:} TQL is a multi-agent distributed solution. As such, learner gNBs are considered the TQL's agents. It is worth mentioning that gNBs are non-cooperative (i.e., they do not share information among themselves). 

    \item \textbf{States:} The state, $s_{l}$, of $l^{th}$ learner gNB is equivalent to the state of an expert gNB as
        \begin{equation}
            s_{l} = 
            \begin{cases}
                s_0, \tabmore \overline{\Gamma}_l \geq \Gamma_{th}, \\
                s_1, \tabmore \text{otherwise},
            \end{cases}
            \label{eq:learnerstates}
        \end{equation}
    where $\overline{\Gamma}_l$ is the average SINR of the $l^{th}$ learner gNB. 

    \item \textbf{Actions:} The actions of the learner is extended to consider joint user-cell association and selection of number of beams (clusters). Indeed, the selection of the number of beams plays a key role in balancing inter-beam and intra-beam interference. On one hand, increasing the number of beams enhances the coverage of users, with less users per beam, which can improve the performance of SIC. On the other hand, more beams leads to higher inter-beam interference, which degrades the performance of decoding at the UE side. Therefore, a gNB should seek to find an optimal number of beams to cover its users. As such, joint user-association and selection of number of beams contribute to increased sum rate per gNB. The actions are formulated as $\boldsymbol{a}_l = [\boldsymbol{\delta}_{u,l}, k_l; u \in \text{\Fontauri\bfseries U}_l, l \in \text{\Fontauri\bfseries L}]$, where $\boldsymbol{\delta}_{u,l}$ represents a vector of logical indicators of $u^{th}$ UE's association to $l^{th}$ learner gNB, and $k_l$ is the number of beams selected by the $l^{th}$ gNB.  

    \item \textbf{Reward:} The learner's reward is equivalent to the expert's reward as in (\ref{eq:reward}) (i.e., the ultimate goal of both the expert and the learner is to improve the average SINR). 

    \item \textbf{Transfer function} The transfer function is used to map a Q-value of a source task to a corresponding Q-value of a target task as shown in Fig. \ref{fig:TvITM}. The transfer process is performed as follows. The learner gNB observes its target state-action pair $(s_t, a_t)$ which is mapped to a source state-action pair $(s_s, a_s)$ using the mapping functions $\phi_s$ and $\phi_a$. With the source state-action pair, the learner gNB addresses the expert's Q-table, stored at the learner gNB, to extract a source Q-value $Q_s(s_s, a_s)$. Afterwards, the source Q-value is mapped to a target Q-value $Q_t(s_t, a_t)$ via a mapping function $\phi_q$. The final Q-value of the learner gNB is represented as follows:
        \begin{equation}
            Q(s_t, a_t) = Q_{t}(s_t, a_t) + Q_{l}(s_t, a_t),
            \label{eq:combinedqvalue}
        \end{equation}
    where $Q_{l}(s_t, a_t)$ is the local Q-value of the learner computed through reinforcement learning as follows:
        \begin{dmath}
            Q_{l}(s_t, a_t)\gets Q_{l}(s_t, a_t) + \alpha [r_l(s_t, a_t) +\gamma \max\limits_{a' \in A} Q_{l}(s_t', a') - Q_{l}(s_t, a_t)],
            \label{eq:qvalueLearner}
        \end{dmath}
    where $r_l(s_t, a_t)$ is the instantaneous reward of the learner gNB, $\alpha$ is a learning rate, $\gamma$ is a discount factor, and $Q_{l}(s_t', a')$ is the expected Q-value at next state $s_t'$. From (\ref{eq:states}) and (\ref{eq:learnerstates}), the expert's (source's) and learner's (target's) states are equivalent ($s_s \equiv s_t$), hence the mapping function of the state is $\phi_s = 1$. In addition, we select the mapping function of the Q-value as $\phi_q = 1$.

    On the other hand, the action's mapping function $\phi_a$ is used to map a target action to a source action. We design the action's mapping function based on inter-beam and intra-beam interference where actions of the learner gNB can be classified into three classes according to the interference level they incur: actions that cause intra-beam interference, actions that cause inter-beam interference, and actions that cause both intra-beam and inter-beam interference. Similarly, actions of the expert gNB can be classified into three classes: actions that cause intra-cell interference, actions that cause inter-cell interference, and actions that cause both. Table \ref{tab:mappingActions} presents the actions' mapping function along with an example on each interference case, where network is comprised of two expert gNBs covering two users and two learner gNBs equipped with up to three beam capability and covering three users. In the first example (i.e., $1^{st}$ row), learner (1) selects one beam to cover the three associated users, which means that learner (2) does not cover any users. As such, learner (1) incurs intra-beam interference only, which maps to a case in which an expert incurs intra-cell interference only (i.e., expert (1) covers all users). In row 2, learner (1) decides to use two beams, hence both inter- and intra-beam interference exist. This action should be mapped to an expert's action that incurs inter- and intra-cell interference. That is, expert (1) covers one out of two users. Without loss of generality, this transfer function can be extended to larger number of users and learner agents.
\end{itemize}
\begin{table}[t]
    \centering
    \caption{Actions' mapping function $\phi_a$. Besides an example of actions' mapping for two expert gNBs with two users and two learner gNBs with three users. Number of beams ranges from one to three.}
    \begin{tabular}{|c|c|c|c|}
         \hline
         \multicolumn{2}{|c|}{\textbf{Interference caused by actions of}} & \multicolumn{2}{c|}{\textbf{Examples}} \\
         \hline
         \multirow{2}{*}{\textbf{Learner-1}} & \multirow{2}{*}{\textbf{Expert-1}} & \textbf{Learner-1} & \textbf{Expert-1} \\
          & & $[\boldsymbol{\delta}_{u,1}, k_1]$ & $[\boldsymbol{\delta}_{u,1}]$ \\
         \hline
         intra-beam & intra-cell & [1 1 1 1] & [1 1]\\
         \hline
         inter-beam  & inter-cell & [1 1 1 3] & [1 0] \\
         \hline
         inter- and intra-beam  & inter- and intra-cell & [1 1 1 2] & [1 0]  \\
         \hline
    \end{tabular}
    \label{tab:mappingActions}
\end{table}
\subsection{Reinforcement Learning}
The formulation of Q-learning is similar to Q-learning of the expert gNB as discussed in section \ref{sec:expertQlearning}, however, the actions are modified to account for joint user-cell association and number of beams selection (i.e., $\boldsymbol{a}_l = [\boldsymbol{\delta}_{u,l}, k_l; u \in \text{\Fontauri\bfseries U}_l, l \in \text{\Fontauri\bfseries L}]$).
\subsection{Best SINR with DBSCAN (BSDC)}
BSDC  performs disjoint user-cell association and cluster, in which user-association is performed based on best SINR (i.e., users associate with the gNB with the best downlink SINR) and DBSCAN is used for clustering (i.e., number of beams selection). DBSCAN is a well known unsupervised learning technique and the details can be found in \cite{3001507}. User clustering has been proposed before in \cite{8454272} using k-means clustering algorithm. In this paper, however, we select DBSCAN for two reasons. First, the main idea of DBSCAN is to discover the points that are closely packed and mark other data points as noise. These closely packed points (or dense distributions) are clusters. Therefore DBSCAN identifies clustered users better in the presence of noise and in node distributions that form non-convex clusters \cite{3001507}. Second, while k-means requires the adjustment of the parameter $k$ (i.e., number of clusters), DBSCAN is able to infer the number of clusters from the given users distribution. 

\subsection{Heuristic Baseline Algorithm:}
To compare the performance of the machine learning algorithms to a simple scheme we include a heuristic algorithm that is considered as our baseline. Similar to BSDC algorithm, the baseline algorithm performs user-cell association and selection of number of beams in a disjoint approach. In particular, user-cell association is performed based on best SINR, in which users associate with the gNB with the best downlink SINR. In addition, clustering (or selection of beams) is performed using a traditional sectorized cell, with fixed number of sectors. In our simulation, we select three sectors. After user-cell association is performed, beam selection is performed based on user's location, i.e., the user is associated with the beam (sector) that covers its location.

\section{Performance Evaluation}\label{sec:perfEval}

\subsection{Simulation Settings}
We use Matlab 5G toolbox to implement a discrete-event simulator, where physical and medium access layers specifications are considered. The simulation parameters of the network model, TQL, Q-learning, and BSDC are presented in Table \ref{tab:simSettings}. In particular, we consider a network with two expert gNBs and two learner gNBs. In case of TQL, expert gNBs perform conventional Q-learning for user-cell association, whereas learner gNBs perform TQL for joint user-cell association and selection of number of beams. In addition, the knowledge (i.e., Q-table) at the expert gNB are transferred to the learner gNBs. In case of Q-learning and BSDC, we do not consider expert gNBs, hence learner gNBs become the only gNBs of the system model (i.e., Fig. \ref{fig:networkModel}b). All gNBs use subcarrier spacing of $15$ KHz and TTI size of $2$ OFDM symbols. Furthermore, link adaptation is performed in conjunction with HARQ technique, where $6$ HARQ processes and a maximum of one HARQ re-transmission were used. All gNBs apply power domain NOMA, which imply that all users are allocated the entire 5G resource block grid. As such, user-cell association controls the load handled by each gNB. Finally, an entire simulation run consumes $6000$ TTIs, whereas $40$ runs are performed to maintain a statistically valid results with confidence interval of $95\%$.
\begin{table}
    \centering
    \caption{5G mm-Wave. Network Simulation Settings}
    \begin{tabular}{|l|l|}
         \hline
         \textbf{\underline{5G-NR}} & \\
         Bandwidth & $20$ MHz \\
         Carrier frequency & $30$ GHz \cite{8782638}\\
         Subcarrier spacing & $15$ KHz \\
         Subcarriers per resource block & $12$ \\
         TTI size & $2$ OFDM symbols  \\
         Max transmission power & $28$ dBm \\
         \hline
         \textbf{\underline{HARQ}} & \\
         Type & Asynchronous HARQ \\
         Round trip delay & $4$ TTIs \cite{8538471} \\
         Number of processes & $6$ \\
         Max. number of re-transmission & $1$ \\
         \hline
         \textbf{\underline{Distribution of users}} & \\
         Distribution & Poisson Cluster Process \\
         Number of users per gNB (Expert) & $3$ \\
         Number of clusters (Expert) & $1$ \\
         Number of users per cluster (Learner) & $6$ \\
         Number of clusters (Learner) & $2$ \\
         Radius of cluster & $30$ m \\
         Total number of users & $18$ \\
         Number of expert gNBs & $2$ \\
         Number of learner gNBs & $2$ \\
         Inter-gNBs distance & $150$ m \cite{8782638} \\
         \hline
         \textbf{\underline{Traffic}} & \\
         Distribution & Poisson \\
         Packet size & $32$ Bytes \\
         \hline
         \textbf{\underline{Q-learning and TQL}} & \\
         Learning rate $(\alpha)$ & $0.5$ \\
         Discount factor $(\gamma)$ & $0.9$ \\
         Exploration probability $(\epsilon)$ & $0.05$ \\
         Threshold SINR ($\Gamma_{th})$ & $20$ dB \cite{6777898,8974494} \\
         \hline
         \textbf{\underline{BSDC}} & \\
         Minimum number of points (minpts) & $1$ \\
         $\epsilon_{BSDC}$ & $40$ \\
         \hline
         \textbf{\underline{Simulation parameters}} & \\
         Simulation time & $6000$ TTI \\
         Number of runs & $40$ \\
         Confidence interval & $95\%$ \\
         \hline
    \end{tabular}
    \label{tab:simSettings}
\end{table}
\subsection{Performance Results I: Complexity and Convergence Analysis}
In this subsection, we provide the complexity and convergence analysis of the proposed algorithms. Complexity analysis considers both runtime and memory complexity. In particular, runtime complexity is presented in Big-O notation computed per gNB per TTI. In addition, memory complexity is presented in number of memory entries required to store information of each algorithm.

Algorithm \ref{alg:algorithmDesign} presents the steps of both Q-learning and TQL approaches. The runtime complexity of the proposed algorithms is presented in Table \ref{tab:Comp}. In particular, complexity of Q-learning stems from two search-for-maximum operations as presented in equations  (\ref{eq:qvalue}) and (\ref{eq:epsilonGreedy}). When using binary search, the complexity of Q-learning becomes $O(\log(2^N)) = O(N)$, where $N$ represents number of users. Similarly, since state, action, and Q-value mapping functions are less complex, a search-for-maximum operation dominates the complexity of TQL. Therefore, complexity of TQL is equivalent to Q-learning (i.e., $O(N)$). As such, Q-learning-based algorithms always incur a runtime complexity in the order of a search-for-maximum operation. On the other hand, the complexity of BSDC is dominated by the DBSCAN algorithm, which is $O(\log(N^2)) = O(\log(N))$ \cite{dbScanComplexity} under binary search assumption. Therefore, BSDC outperforms Q-learning and TQL in runtime complexity.
\begin{algorithm}[t]
    \begin{algorithmic}[1]
    \FOR{scheduling assignment period $t$ = 1 to $T$}
          \STATE \underline{\textbf{Step 1:}} gNB receives feedback from users in the form of SINRs. 
          \STATE \underline{\textbf{Step 2:}} Observe next state as in (\ref{eq:states}) for Q-learning or (\ref{eq:learnerstates}) for TQL.
          \STATE \underline{\textbf{Step 3:}} Update Q-value as in (\ref{eq:qvalue}) for Q-learning or (\ref{eq:qvalueLearner}) for TQL.
          \STATE \underline{\textbf{Step 4:}} Select action through $\epsilon$-greedy approach.
          \begin{equation}
              \boldsymbol{a} = 
              \begin{cases}
              \text{Random}, \tabmore \tabmore \tabmore \tabmore \tabmore (1-\epsilon), \\
              \arg \max\limits_{a'} Q(s, a'), \tabmore \epsilon.
              \end{cases}
              \label{eq:epsilonGreedy}
          \end{equation}
    \ENDFOR
    \end{algorithmic}
    \caption{Q-learning and TQL algorithms for user-cell association and selection of number of beams}
    \label{alg:algorithmDesign}
\end{algorithm}
On the other hand, space complexity is presented in Table \ref{tab:Comp}. In particular, Q-learning with its tabular version requires a Q-table of size $nStates \times nActions$. Therefore, the space complexity of Q-learning becomes $O(2 \times (K \times 2^N) )= O(K \times 2^N)$, where $K$ represents the possible values of number of beams. Similarly, TQL requires two Q-tables, one for its local Q-learning, whereas the other one is the transferred Q-table from the expert gNB. In particular, the space complexity of the local Q-table is $O(K \times 2^N)$, whereas the complexity of the transferred Q-table's is $O(2^N)$ since expert gNB does not consider the selection of number of beams. As such, the total space complexity of TQL becomes $O(K \times 2^N)$. Finally, BSDC is dominated by the memory requirement of DBSCAN, which is $O(N)$ \cite{dbScanComplexity}. To summarize, BSDC outperforms both Q-learning and TQL with respect to both runtime and space complexity. However, it is worth mentioning that space complexity of Q-learning, and TQL, can be reduced by employing deep Q-learning as proposed in \cite{8647289}, where deep Q-learning replaces the need for a Q-table by directly predicting Q-values using a deep neural network.
\begin{table}
	\centering
	\caption{Complexity comparison among the proposed algorithms}
    \label{tab:Comp}
		\begin{tabular}{|c|c|c|c|}
		\hline
        \textbf{Complexity} & \textbf{Q-learning} & \textbf{TQL} &  \textbf{BSDC} \\
        \hline
        \textbf{Runtime} & $O(N)$ & $O(N)$ & $O(\log(N))$ \\
        \hline
        \textbf{Space} & $O(K \times 2^N)$ & $O(K \times 2^N)$ & $O(N)$ \\
        \hline
        \end{tabular}
\end{table}

\begin{figure}
    \centering
    \includegraphics[scale=0.25]{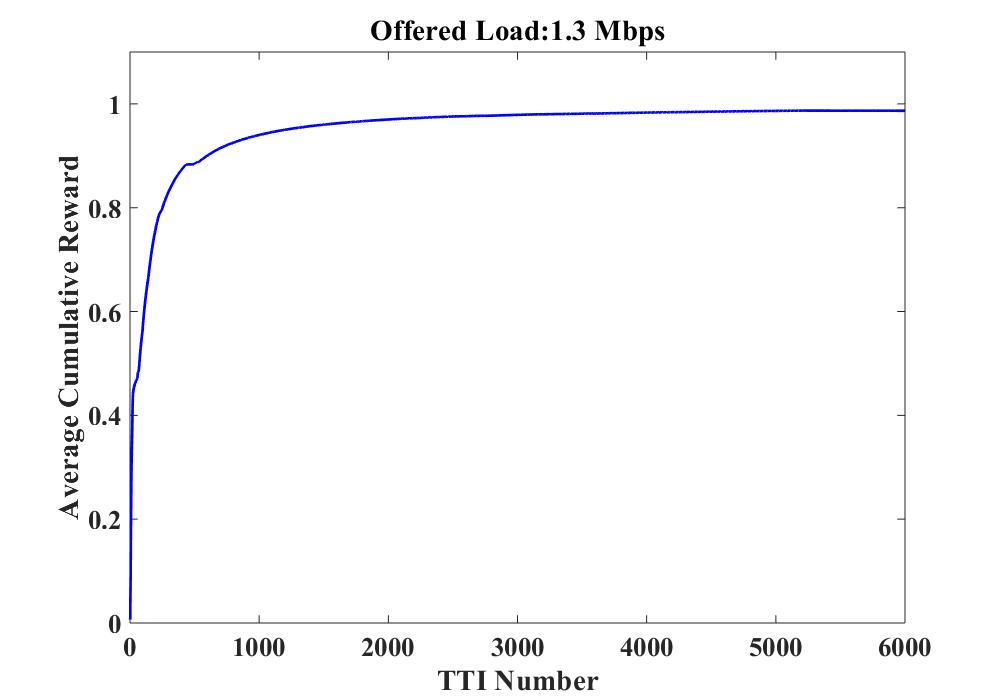}
    \caption{Convergence of expert gNBs represented by the average cumulative reward.}
    \label{fig:convExpert}
\end{figure}

\begin{figure}
    \centering
    \includegraphics[scale=0.25]{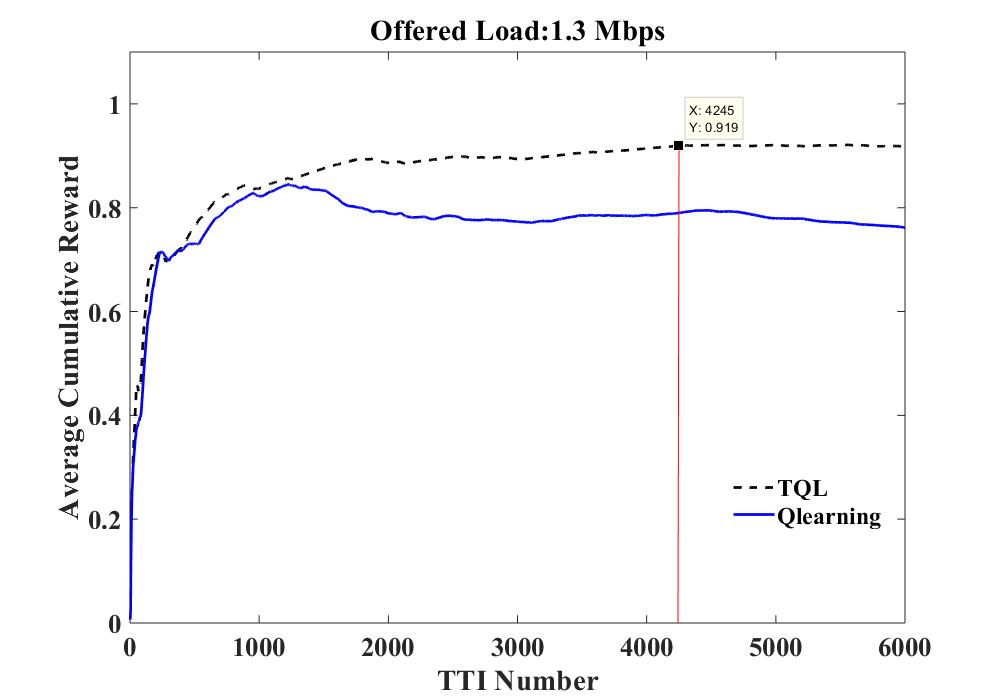}
    \caption{Convergence of learner gNBs represented by the average cumulative reward. Total offered load is $1.3$ Mbps.}
    \label{fig:convergence}
\end{figure}

The convergence of the expert gNB is presented in Fig. \ref{fig:convExpert}. Note that in case of TQL, the expert gNB is performing user-cell association solely. In the figure, the average cumulative reward is plotted against iteration number (i.e., TTI number). The figure demonstrates the successful convergence of the expert agent within the lifetime of the simulation (i.e., $6000$ TTIs). This is essential since results of expert gNB beyond convergence is transferred to the learner gNB. The convergence of learner gNBs for both Q-learning and TQL is plotted in Fig. \ref{fig:convergence}. As observed from the figure, TQL outperforms Q-learning in two aspects. First, TQL converges rapidly (i.e., around $4245$ TTIs), whereas Q-learning experiences more iterations with a sign of convergence toward the end of the simulation time. Although we select TTI length of $2$ OFDM symbols, the convergence trend will not be impacted by the choice of other TTI configurations of 5G-NR. However, the time for convergence will be longer and proportional to the length of TTI duration in ms. Second, TQL achieves higher cumulative reward, whereas Q-learning dwells around lower cumulative reward. This demonstrates TQL's ability to converge to a better policy for user-cell association and selection of number of beams. This constitutes an advantage for TQL compared to Q-learning. While both have the same complexity as shown in table \ref{tab:Comp}, TQL converges faster than Q-learning. It is also worth mentioning that the transferred Q-table from the expert can be learned in an offline setup. This means that TQL can train experts offline whereas learners can be trained in the field with an online algorithm in a shorter time.
\subsection{Performance Results II: Stationary Users}
In this subsection, simulation results are provided for the proposed algorithms under stationary users scenario (i.e., no mobility). In particular, PCP is used for initial positions of users. 

\begin{figure}
    \centering
    \includegraphics[scale=0.25]{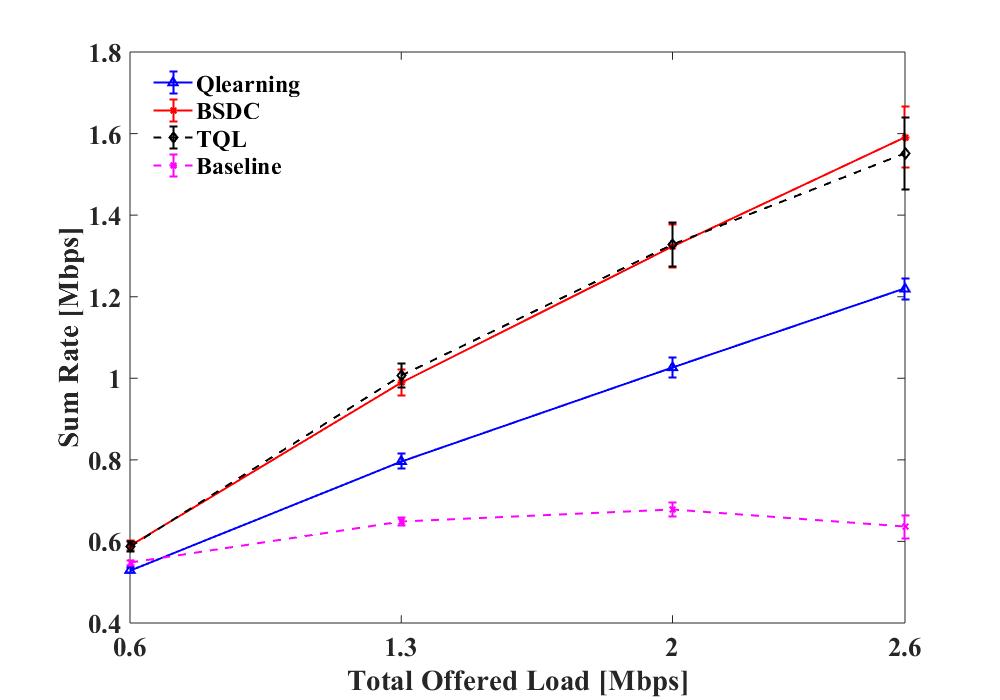}
    \caption{Sum rate in [Mbps] of learner gNBs against total offered network load in [Mbps] under PCP deployment of users.}
    \label{fig:rate}
\end{figure}

\begin{figure}
    \centering
    \includegraphics[scale=0.25]{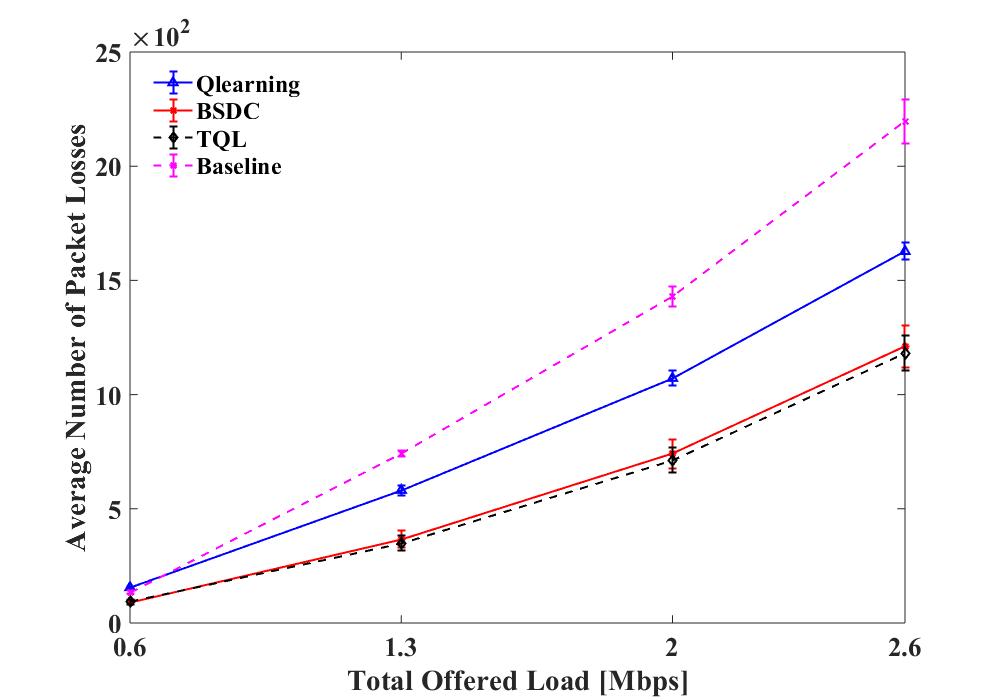}
    \caption{Average number of packet loss in [packets] against total offered network load in [Mbps] under PCP deployment of users.}
    \label{fig:pdr}
\end{figure}

\begin{figure}
    \centering
    \includegraphics[scale=0.25]{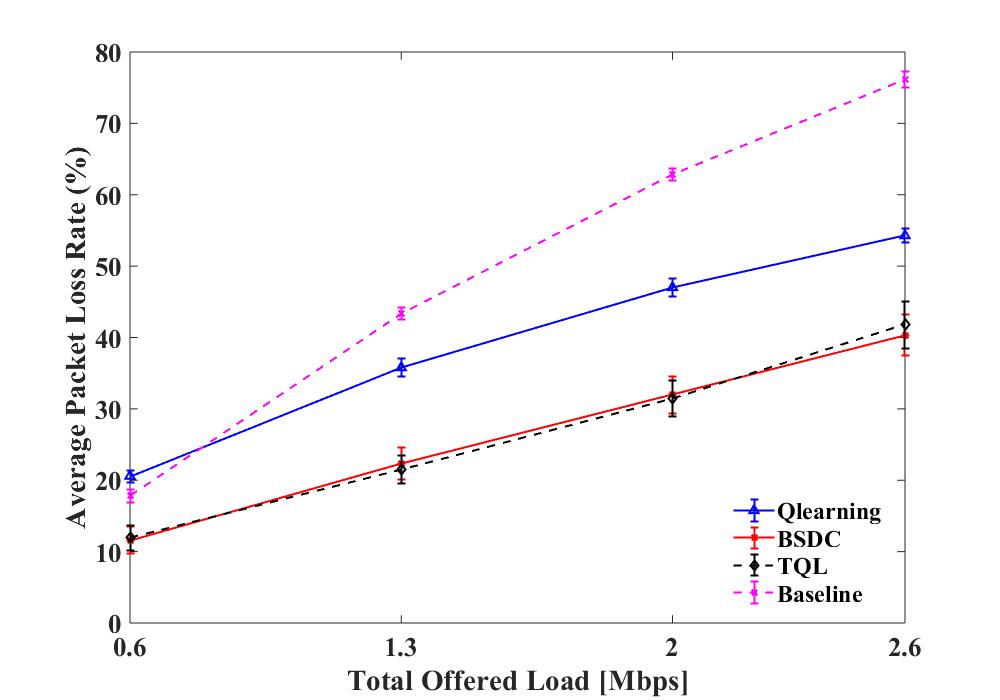}
    \caption{Packet loss rate in [\%] against total offered network load in [Mbps] under PCP deployment of users.}
    \label{fig:pdrpercent}
\end{figure}

As presented in Fig. \ref{fig:convergence}, TQL proved to converge to a better policy for user-cell association and selection of number of beams. This is evident from Fig. \ref{fig:rate}, where the sum rate of the learner gNBs is plotted against the total offered load in the network. In particular, TQL outperforms Q-learning under all traffic loads with about $23\%$ improvement at the highest traffic load. BSDC, on the other hand, performs very closely to TQL. This was expected since BSDC uses DBSCAN clustering which performs very well under the PCP distribution model. In the next subsection, we perform comparison under different user distribution and mobility model to highlight the superiority of TQL compared to DBSCAN. 
Besides achieving high rate, the rate of TQL is close to the total offered rate in the network, which implies high reliability as well. This is highlighted in Fig. \ref{fig:pdr} and in Fig. \ref{fig:pdrpercent} which plot the the packet loss against the total offered traffic load. The figures demonstrate that TQL outperforms Q-learning under all traffic conditions. In addition, all three machine learning algorithms outperform the baseline algorithm as shown in Fig. \ref{fig:rate}, Fig. \ref{fig:pdr}, and Fig. \ref{fig:pdrpercent}. 

\begin{figure}[t]
    \centering
    \includegraphics[scale=0.25]{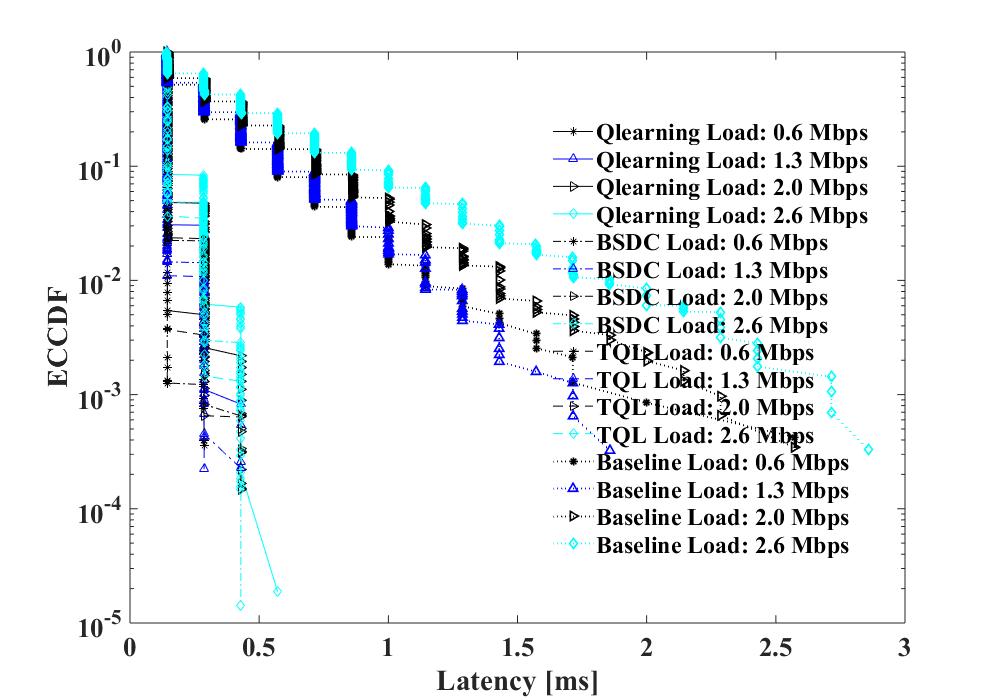}
    \caption{Empirical Complementary Cumulative Distribution Function (ECCDF) of latency for different total offered network load under PCP deployment of users.}
    \label{fig:latency}
\end{figure}

Finally, Fig. \ref{fig:latency} presents the empirical Complementary Cumulative Distribution Function (ECCDF) of latency, where latency is defined as the end-to-end delay of successfully received packets from gNB to users. The figure shows close performance of the three proposed algorithms. This was expected since all algorithms do not account for latency improvement (Refer to the reward function in (\ref{eq:reward})). Furthermore, the low latency achieved (i.e., latency below $1$ msec) is due to the restriction put on the number of HARQ re-transmissions, where one re-transmission is used in our simulation \cite{8361404, 9014032}. On the other hand, the baseline algorithm incurs higher latency compared to the three proposed algorithms.

\subsection{Performance Results III: Random Waypoint Mobility}
The mobility of users might have a significant impact on the performance of the proposed algorithms. Mobile users tend to change their clustering behavior which might lead to different number of clusters with iterations. This mandates rapid response in terms of number of beams selection. Furthermore, due to mobility, users might change the clusters they belong to, which also impacts user-cell association. As such, enhancing performance under mobility becomes a necessary component of the learning algorithm. In this subsection, we assess the performance of the proposed algorithms under a random waypoint mobility scenario. In particular, initial users' deployment follows PCP distribution whereas mobility of users follows random waypoint model. 

\begin{figure}
    \centering
    \includegraphics[scale=0.25]{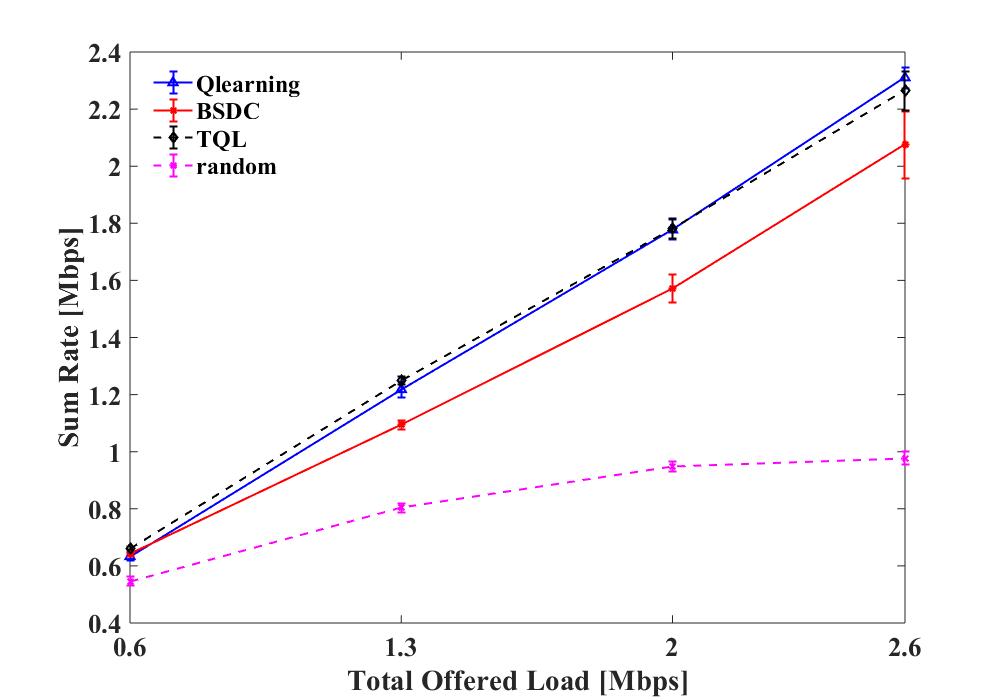}
    \caption{Sum rate in [Mbps] of learner gNBs against total offered network load in [Mbps] under random waypoint mobility of users.}
    \label{fig:rateMobility}
\end{figure}

\begin{figure}[t]
    \centering
    \includegraphics[scale=0.25]{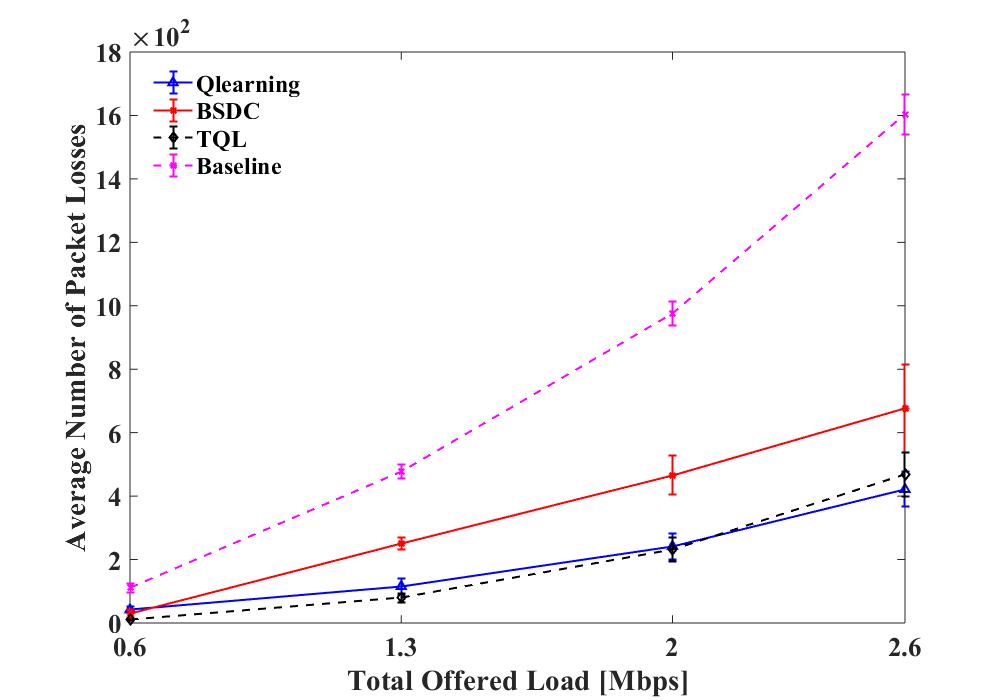}
    \caption{Average number of packet loss in [packets] against total offered network load in [Mbps] under random waypoint mobility of users.}
    \label{fig:pdrMobility}
\end{figure}

\begin{figure}
    \centering
    \includegraphics[scale=0.25]{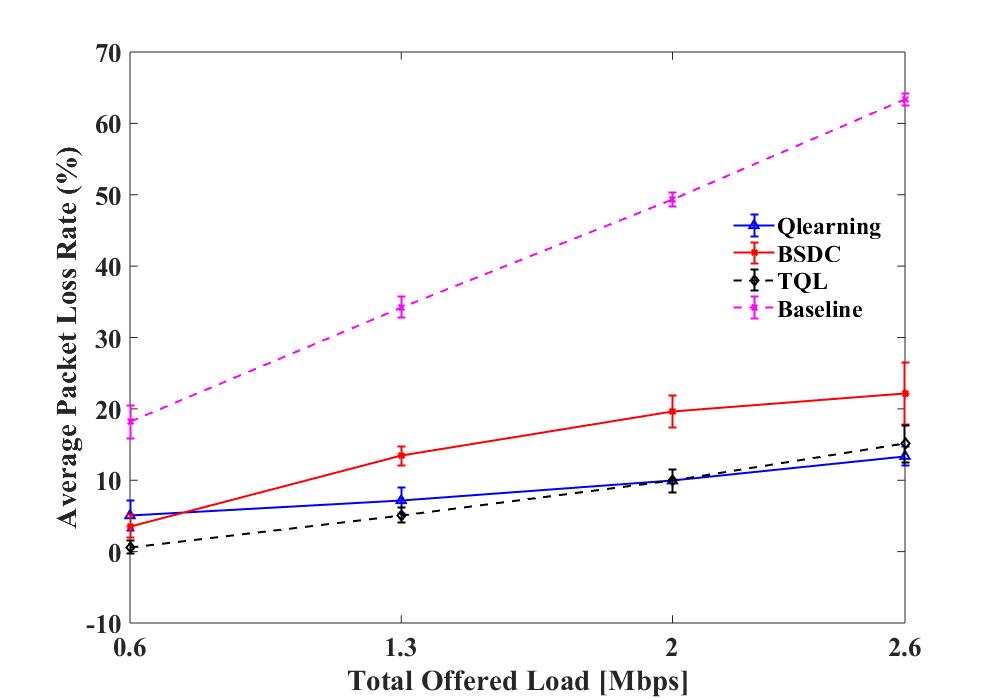}
    \caption{Packet loss rate in [\%] against total offered network load in [Mbps] under random waypoint mobility of users.}
    \label{fig:pdrMobilitypercent}
\end{figure}

Fig. \ref{fig:rateMobility}, Fig. \ref{fig:pdrMobility}, and Fig. \ref{fig:pdrMobilitypercent} present the sum rate of learner gNBs, number of packet loss, and packet loss rate in percentage versus total offered load under random waypoint mobility model, respectively. Unlike stationary case, TQL and Q-learning outperform BSDC performance in both sum rate and packet loss. In particular, TQL and Q-learning demonstrate $12\%$ sum rate improvement over BSDC at the highest offered traffic load. Again, the machine learning algorithms outperform the baseline algorithm.

\section{Conclusion}\label{sec:conclusion}
In this work, we presented three machine learning algorithms for joint user-cell association and selection of number of beams in mm-Wave networks for the purpose of sum rate maximization. The first algorithm is a transfer reinforcement learning algorithm that aims at improving the convergence of reinforcement learning through knowledge transfer from an expert agent to a learner agent. To the best of our knowledge, this is the first time that a transfer reinforcement learning algorithm is proposed for optimizing sum rate in mm-Wave networks. The expert agent performs the simple task of user-cell association using Q-learning, whereas the learner agent aims to build upon the expert's knowledge and performs the complex task of joint user-cell association and selection of number of beams to cover the associated users. The second algorithm is conventional Q-learning that performs joint user-cell association and selection of number of beams. The third algorithm is a combination of best SINR for user-cell association and DBSCAN for users clustering technique. The results demonstrate the suitability of each algorithm to the deployment scenarios considered. Under mobility scenario, TQL and Q-learning demonstrate $12\%$ sum rate improvement over BSDC at the highest offered traffic load, whereas under stationary scenario, Q-learning and BSDC outperforms TQL with about $10-23\%$ at lowest and highest offered traffic loads, respectively. In addition, BSDC has lower complexity than the other techniques and TQL has faster convergence than the Q-learning based technique. Besides, TQL offers a unique advantage for offline learning of a task and transferring the knowledge to another task with online learning in the field. 

\section*{Acknowledgment}
This research is supported by the Natural Sciences and Engineering Research Council of Canada (NSERC) Canada Research Chairs program.

\ifCLASSOPTIONcaptionsoff
  \newpage
\fi

\bibliographystyle{IEEEtran.bst}
\bibliography{main_201206}

\begin{IEEEbiography}[{\includegraphics[width=1in,height=1.2in,clip,keepaspectratio]{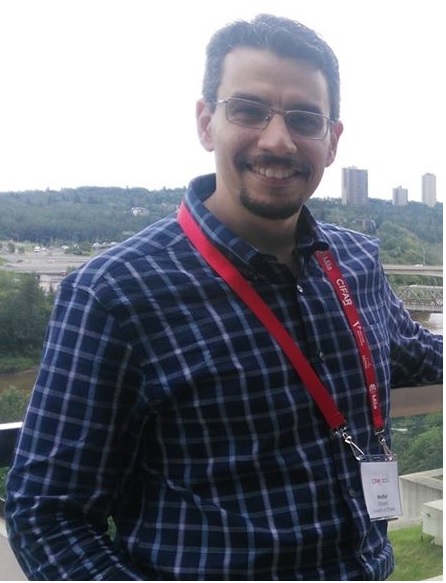}}]{Medhat Elsayed} is a PhD candidate at the University of Ottawa. He obtained his BSc and MSc degrees from Cairo University, Egypt, in 2009 and 2013 respectively. He gained extensive research experience in optimizing the performance of wireless systems while at Cairo University, Qatar University and University of Ottawa. His current research interests focus on developing innovative machine learning algorithms for 5G wireless networks and beyond. Mr. Elsayed has received the International Doctoral Scholarship from the University of Ottawa in 2018. He also received the NSERC CREATE TOPSET graduate research training scholarship from the University of Ottawa in 2018. Mr. Elsayed delivered a tutorial at the Second Annual Workshop of the Ottawa AI Alliance held in NRC of Ottawa in 2019. The tutorial covered background on deep learning and reinforcement learning, a survey on the state-of-art work, and his recent work in the field. He also collaborated with a poster presentation at the Deep Learning and Reinforcement Learning Summer School held in the University of Alberta in 2019. The poster covered his work on using reinforcement learning for ultra-reliable low-latency communication.  He is the author of several publications addressing wireless communication problems via machine learning techniques. In addition, he is a co-inventor of a filed patent.
\end{IEEEbiography}

\begin{IEEEbiography}[{\includegraphics[width=1in,height=1.2in,clip,keepaspectratio]{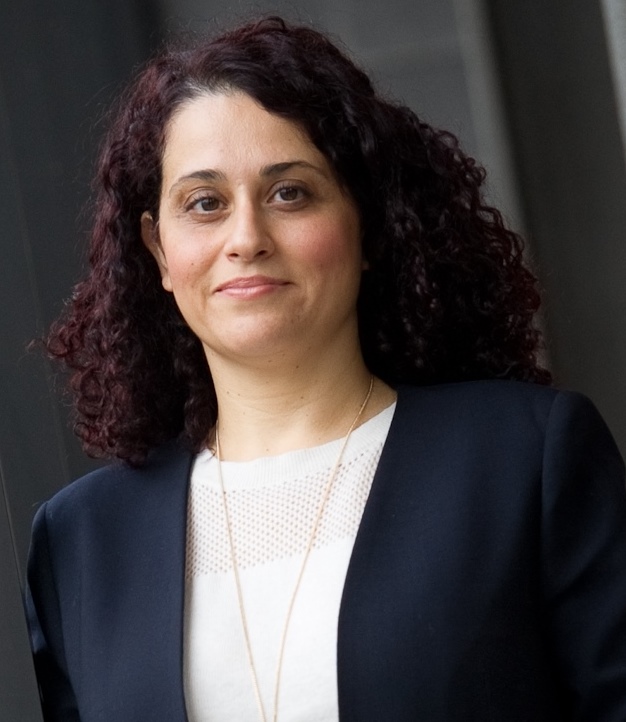}}]{Melike Erol-Kantarci} is Tier 2 Canada Research Chair in AI-enabled Next-Generation Wireless Networks and Associate Professor at the School of Electrical Engineering and Computer Science at the University of Ottawa. She is the founding director of the Networked Systems and Communications Research (NETCORE) laboratory. She has received awards and recognitions. Dr. Erol-Kantarci is the co-editor of three books on smart grids, smart cities and intelligent transportation. She has delivered 50+ keynotes, plenary talks and tutorials around the globe. She is on the editorial board of the IEEE Transactions on Cognitive Communications and Networking, IEEE Internet of Things Journal, IEEE Communications Letters, IEEE Networking Letters, IEEE Vehicular Technology Magazine and IEEE Access. She has acted as the general chair and technical program chair for many international conferences and workshops. Her main research interests are AI-enabled wireless networks, 5G and 6G wireless communications, smart grid, Internet of things and wireless sensor networks. She is a senior member of the IEEE and the ACM. 
\end{IEEEbiography}

\begin{IEEEbiography}[{\includegraphics[width=1in,height=1.2in,clip,keepaspectratio]{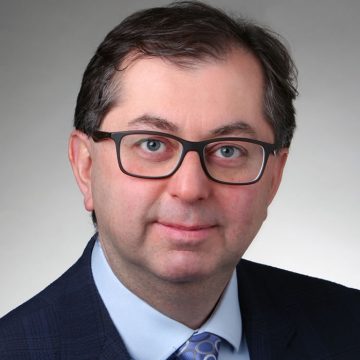}}]{Halim Yanikomeroglu}
is a full professor in the Department of Systems and Computer Engineering at Carleton University. His research interests cover many aspects of wireless technologies with special emphasis on wireless networks. His collaborative research with industry has resulted in $36$ granted patents. He is a Fellow of IEEE, a Fellow of the Engineering Institute of Canada (EIC) and the Canadian Academy of Engineering (CAE). He is a Distinguished Speaker for both the IEEE Communications Society and the IEEE Vehicular Technology Society.
\end{IEEEbiography}




\end{document}